\documentclass[a4paper,11pt]{article}
\pdfoutput=1 % if your are submitting a pdflatex (i.e. if you have
\usepackage{jcappub} % for details on the use of the package, please
\usepackage[T1]{fontenc} % if needed
\usepackage{graphicx}

\def\aj{AJ}

\def\mnras{MNRAS}

\def\nat{Nature}

\title {\boldmath Galaxy interactions in different environments: An
  analysis of galaxy pairs from the SDSS}

\author[a]{Apashanka Das,}

\author[a]{Biswajit Pandey,}

\author[b]{Suman Sarkar,}

\author[a]{and Arunima Dutta}
\affiliation[a]{Department of Physics, Visva-Bharati University,
  Santiniketan, 731235, India}
\affiliation[b]{Department of Physics, Indian Institute of Science
  Education and Research Tirupati, Tirupati - 517507, Andhra Pradesh,
  India} 
\emailAdd{a.das.cosmo@gmail.com}
\emailAdd{biswap@visva-bharati.ac.in}
\emailAdd{suman2reach@gmail.com}
\emailAdd{aru.megha@gmail.com}

\abstract{We analyze the galaxy pairs in a volume limited sample ($M_r
  \leq -21$) from the SDSS to study the effects of galaxy interactions
  on the star formation rate and colour of galaxies in different
  environments. We study the star formation rate and colour of the
  paired galaxies as a function of projected separation and compare
  the results with their control samples matched in stellar mass,
  redshift and local density. We find that the major interactions
  significantly enhance the star formation rate in paired galaxies and
  turn them bluer with decreasing pair separation within $30$ kpc. The
  impact of tidal interactions on star formation rate and colour are
  more significant in the heavier members of the major pairs. The star
  formation enhancement in major pairs is significantly higher at the
  low-density environments, where the influence can extend up to $\sim
  100$ kpc. Contrarily, the major pairs at high-density environments
  show suppression in their star formation. Depending on the embedding
  environments, the major interactions in the intrinsically brighter
  galaxy pairs can thus enhance or quench star formation. We find that
  the minor pairs at both low-density and high-density environments
  are significantly less star-forming and redder than their control
  galaxies. It indicates that the minor interactions in intrinsically
  brighter galaxy pairs always suppress the star formation
  irrespective of their environment. The lighter members in these
  minor pairs show a greater susceptibility to suppressed star
  formation. Our results imply that both the major and minor
  interactions can contribute to the observed bimodality. We conclude
  that the galaxy evolution is determined by a complex interplay
  between the galaxy properties, galaxy interactions, and
  environment.}

\begin{document}
\maketitle
\flushbottom

%\begin{keywords}
%methods: statistical - data analysis - galaxies: formation - evolution
%- cosmology: large scale structure of the Universe.
%\end{keywords}

\section{Introduction}

Understanding the origin of the galaxies and their evolution is one of
the most challenging goals of modern cosmology. It remains one of the
most active and fertile areas of research in the last few decades. In
the current paradigm, the primordial density fluctuations in the dark
matter density field collapse into dark matter halos, forming the
first bound objects. These dark matter halos accrete neutral hydrogen
gas from their surrounding environment that eventually leads to the
formation of galaxies at their centres by the cooling and condensation
\citep{reesostriker77, silk77, white78, fall80}. In this picture, the
gas accretion from the intergalactic medium (IGM) is the primary
mechanism responsible for the growth of a galaxy. The subsequent
formation of a supermassive black hole at the centre of the galaxy and
an efficient accretion onto it can trigger AGN activity. The jet from
the AGN, starburst winds and supernovae explosions can drive gas
outflows from the galaxy. The gas ejected from the galaxy eventually
cools down and reaccretes again to continue the cycle. The interplay
between these processes and their equilibrium are important in galaxy
evolution. However, the evolutionary history of a galaxy is not
entirely determined by these processes alone.

The observations suggest a significant evolution of galaxy properties
in the recent past. The cosmic star formation declines by an order of
magnitude between $z=1$ to $z=0$ \citep{madau96}. It is now well known
that the number of red galaxies and their stellar mass are steadily
increasing since $z \sim 1$ \citep{bell04b, faber07}. On the other
hand, the stellar mass in the blue galaxies remains nearly the same
during this period \citep{faber07, martin07, ruhland09}. These
observational results can be understood if the blue star-forming
galaxies transform into red galaxies via quenching of star formation,
and the galaxies in the blue cloud experience an enhanced star
formation.

It is important to remember that the galaxies are not island universes
that evolve in isolation. They are an integral part of a more
extensive and complex network, namely the cosmic web
\citep{bond96}. Both initial conditions and interactions with the
small and large scale environment play crucial roles in their
formation and evolution. In the hierarchical scenario, the galaxy
interactions and mergers provide an efficient mechanism for the
buildup of massive galaxies. Such processes can modify the mass
distribution and morphology of galaxies and trigger star formation
activity. Using simulations of tidal interactions, Toomre \& Toomre
\citep{toomre72} first showed that the spiral and irregular galaxies
could transform into ellipticals and S0 galaxies. Subsequent studies
with more sophisticated simulations \citep{barnes96, mihos96,
  tissera02, cox06, dimatteo07, montuori10, rupke10, torrey12,
  renaud14} revealed that the tidal torques generated during the
encounter act upon the misaligned gaseous and stellar bars that allows
the stellar component to remove angular momentum from the gas. The
outward transfer of angular momentum activates gas inflows towards the
centre of the galaxy. The rapid increase in gas density near the
centre triggers a starburst that can efficiently convert the available
gas reservoir into stars. The efficiency of this tidally triggered
star formation is known to depend on several factors such as the
amount of available gas, depth of the potential well, morphology,
orbital parameters and the internal dynamical properties of galaxies
\citep{barnes96, tissera00, perez06}.

The simulations of galaxy interactions suggest that the gravitational
tidal torque generated during the interaction is more potent in galaxy
pairs with similar stellar mass or luminosity. The large-scale gas
inflows resulting from such interactions trigger new star formation in
these galaxies. These interactions are known as the major
interactions. The interactions between galaxies with a larger mass or
luminosity contrast are known as minor interactions. It is well known
that the frequency of dark matter halo mergers increases with the mass
ratio \citep{lacey93, fakhouri08}. So the minor interactions and
mergers are more frequent in a galaxy's history. They are also known
to play an essential role in triggering star formation and growth of
galaxies in the semi-analytic models \citep{somerville01, guo08,
  bell08}. Studies with simulations \citep{mihos94, mastro05, cox08}
indicate that a lower level of star formation enhancement may also
occur in minor mergers after several billion years. The induced star
formation in minor interactions and mergers are known to depend on the
structural and orbital parameters of the galaxies.

The first observational evidence of enhanced star formation in
interacting galaxies came from a seminal study of optical colours in
the morphologically disturbed galaxies by \citep{larson78}. Many
further studies of interacting galaxy pairs from the modern
spectroscopic redshift surveys have now confirmed the SFR enhancement
at smaller pair separation \citep{barton00, lambas03, alonso04,
  nikolic04, alonso06, woods06, woods07, barton07, ellison08,
  heiderman09, knapen09, robaina09, ellison10, woods10, patton11}. The
level of enhancement reported in most of these studies is within a
factor of two compared to the isolated galaxies. The enhancement is
known to depend on multiple factors such as the separation, luminosity
or mass ratio and the type of galaxies involved in the interaction.
The projected separation between the members in a pair is the most
widely used indicator of the merger phase of a galaxy pair. The pairs
at smaller separation are most likely undergoing a close passage,
thereby triggering a starburst. On the other hand, the pairs at larger
separation are believed to be receding away after their first
pericentric passage and hence show a lower efficiency of star
formation. However, these are difficult to confirm as the projected
separation corresponds to a snapshot view of the interaction and does
not provide any direct information about the time scale.

Most of the observational studies of galaxy pairs confirm the tidally
triggered star formation in major interactions. However, the issue of
star formation enhancement in minor interactions are less clear. In
the hierarchical galaxy formation model \citep{somerville99,
  kauffmann1, kauffmann2, diaferio99}, most interactions and mergers
occur between unequal-mass systems due to the greater abundance of low
mass and low luminosity galaxies. The minor interactions may thus play
a crucial role in galaxy evolution. Any observational study of minor
interaction and merger is challenging due to several reasons. The
number of minor pairs identified from magnitude limited surveys are
far less than the number of major pairs as the galaxies have similar
magnitudes in such surveys. It is also difficult to identify the
low-luminosity companions around the more luminous members due to the
contaminations from the background galaxies. Despite these
limitations, the effects of minor interactions on star formation have
drawn considerable interest. Lambas et al. \citep{lambas03} study the
star formation enhancement in paired galaxies using 2dFGRS and find a
dependence on the relative luminosity of the pairs. Nikolic et
al. \citep{nikolic04} use SDSS to analyze the star formation in paired
galaxies and find no dependence on the luminosity of the companion
galaxy. Woods et al. \citep{woods06} analyze data from cfA2 survey and
a follow-up search to find that the star formation enhancement in
pairs decreases with increasing stellar mass ratio. Woods \& Geller
\citep{woods07} show that the specific SFR of the less massive member
in a minor pair is enhanced, whereas the more massive member remains
unaffected.  Ellison et al. \citep{ellison08} analyze the SDSS data
and find tentative evidence for higher SFR for the less massive
companions in minor pairs at a low significance level. Li et
al. \citep{li08} also reached a similar conclusion using the SDSS
data. Scudder et al. \citep{scudder12b} study the SFR enhancements in
SDSS galaxy pairs and find that both major and minor mergers show
significant SFR enhancements. Kaviraj et al. \citep{kavi11} find that
the minor mergers may induce moderate star formation in early-type
galaxies at low redshift. Lambas et al. \citep{lambas12} show that the
minor mergers are roughly two times less efficient in forming new
stars than the major mergers.

The galaxies grow in mass by smooth accretion, accretion from
companions and mergers with other galaxies. The observational
signatures of major interactions and mergers are well understood.  But
our current understanding of the impact of minor interactions and
mergers are far from complete. Observational studies indicate a low
level of star formation enhancement in the less massive members of the
pairs. However, it is unclear if the SFR enhancement in minor
interactions is equally effective at all luminosities. The ratio of
luminosity or mass may not wholly decide the SF activity in minor
pairs. The exact ratio of the stellar mass or luminosity may affect
the induced star formation differently in separate luminosity ranges
and environments. Most of the earlier studies on galaxy pairs
primarily focus on star-forming galaxies. It would be also interesting
to know the effectiveness of tidal interaction in triggering star
formation in the intrinsically more luminous galaxy pairs with
relatively lower star formation.

The environments of galaxies are known to play a driving role in the
formation and evolution of galaxies \citep{oemler74, davis76, dress80,
  guzo97, zevi02, hog03, blan03, einas03, gotto03, kauffmann04,
  pandey06, park07, mocine07, pandey08, porter08, bamford09, cooper10,
  koyama13, pandey17, sarkar20, bhattacharjee20, pandey20}. The star
formation activity is known to get suppressed in the high-density
regions \citep{lewis02,gomez03,kauffmann04}. The environment can
quench star formation in galaxies through ram pressure stripping
\citep{gunn72}, galaxy harassment \citep{moore96, moore98},
strangulation \citep{gunn72, balogh00} and starvation \citep{larson80,
  somerville99, kawata08}. The same is also true for the efficiency of
star formation in galaxy pairs in the denser environments. The tidal
interactions in such environments can cause gas loss through
starburst, AGN or shock-driven winds \citep{cox04, murray05,
  springel05}. Perez et al. \citep{perez09a} find that the red
fraction in galaxy pairs increases at intermediate density that
indicates an efficient pre-processing by close encounters in such
environments. In general, the galaxy pairs inhabit relatively denser
regions compared to the isolated galaxies. It is possible to attribute
the observed differences in the paired and isolated galaxies
simultaneously to the tidal interaction and the environment, leading
us to an age-old problem of ``nature versus nurture'' in galaxy
formation and evolution. Although the control sample of the isolated
galaxies is often constructed by matching the stellar mass and
environment, it would be interesting to know the influence of major
and minor interactions on the colour and star formation activity of
the galaxies in the high and low-density regions of the cosmic density
field. Most of the earlier studies on galaxy interactions analyzed
flux-limited samples. It is difficult to meaningfully define the local
physical environment around galaxy pairs in a flux limited
sample. This can be more reliably estimated in a volume limited
sample. We plan to use a volume limited sample for the present work.

The Sloan Digital Sky Survey (SDSS) \citep{strauss02} is the largest
photometric and spectroscopic redshift survey available at present.
The availability of precise spectroscopic information for a large
number of galaxies in the SDSS provides an excellent opportunity for
the statistical study of galaxy interactions and their effects on the
star formation and colour. The galaxy colour is strongly correlated
with the star formation due to the observed bimodality
\citep{strateva01, baldry04}. The galaxies in the blue cloud are gas
rich and they have higher star formation rates. Contrarily, the red
sequence hosts the gas poor galaxies with very low to no star
formation. The tidal interactions between galaxies may trigger
starbursts and consequently can alter their colours. Such colour
changes usually happen on a time scale longer than the starburst. The
effect of tidal interactions on the galaxy pairs can be captured more
convincingly if we employ both star formation and colour for such
studies. We plan to study the star formation rate and the dust
corrected $(u-r)$ colour of the galaxy pairs in a volume limited
sample from the SDSS, as a function of the projected separation and
then compare them with the same from control samples of the isolated
galaxies. The control samples will be prepared by simultaneously
matching the stellar mass, local density and redshift of the paired
galaxies. We want to study the colour and star formation activity of
the major and minor pairs in both the low-density and high-density
regions that would allow us to understand the relative role of tidal
interaction and the environment in deciding these galaxy properties.

The outline of the paper is as follows. We describe the data and the
method of analysis in Section 2, discuss the results in Section 3 and
present our conclusions in Section 4.

Through out the paper we use the $\Lambda$CDM cosmological model with
$\Omega_{m0}=0.315$, $\Omega_{\Lambda0}=0.685$ and $h=0.674$
\citep{planck18} for conversion of redshifts to comoving distance.

\section{DATA AND METHOD OF ANALYSIS}   

\subsection{SDSS DR16}

The Sloan Digital Sky Survey (SDSS) is a multi-band imaging and
spectroscopic redshift survey with a 2.5 m telescope at Apache Point
Observatory in New Mexico. The technical details of the SDSS telescope
is described in Gunn et al. \citep{gunn06} and a description of the
SDSS photometric camera can be found in Gunn et
al. \citep{gunn98}. The selection algorithm for the SDSS Main galaxy
sample is discussed in Strauss et al. \citep{strauss02} and a
technical summary of the survey is provided in York et
al. \citep{york00}.

We use the sixteenth data release (DR16) of the SDSS to identify the
galaxy pairs in the nearby universe. We use structured query language
(SQL) to download the spectroscopic and photometric information of
galaxies in DR16 \citep{ahumada20} from the SDSS
CasJobs \footnote{https://skyserver.sdss.org/casjobs/}. We select a
contiguous region of the sky spanned by the equatorial coordinates
$135^{\circ} \leq \alpha \leq 225^{\circ}$ and $0^{\circ} \leq \delta
\leq 60^{\circ}$ for our analysis and construct a volume limited
sample by restricting the $r$-band apparent magnitude to $m_r < 17.77$
and the absolute $r$-band Petrosian magnitude to $M_r \leq -21$. The
resulting sample extends up to $z \leq 0.115$ and consists of $92102$
galaxies.

We obtain the stellar mass and the star formation rate (SFR) of the
galaxies from the {\it{StellarMassFSPSGranWideNoDust}} table. These
are calculated using the Flexible Stellar Population Synthesis (FSPS)
model \citep{conroy09}. The effects of the fibre aperture in these
measurements are appropriately taken into account by using the
aperture correction schemes described in Brinchmann et
al. \citep{brinchmann04}. We retrieve the information of internal
reddening E(B-V) of the galaxies from {\it{emmissionLinesPort}} table
which is based on publicly available Gas AND Absorption Line Fitting
(GANDALF) \citep{gandalf} and penalised PIXEL Fitting (pPXF)
\citep{ppxf}. Throughout this analysis, we have used dust corrected
$u-r$ colour of the galaxies.

\begin{figure*}[htbp!]
\centering
\includegraphics[width=7cm]{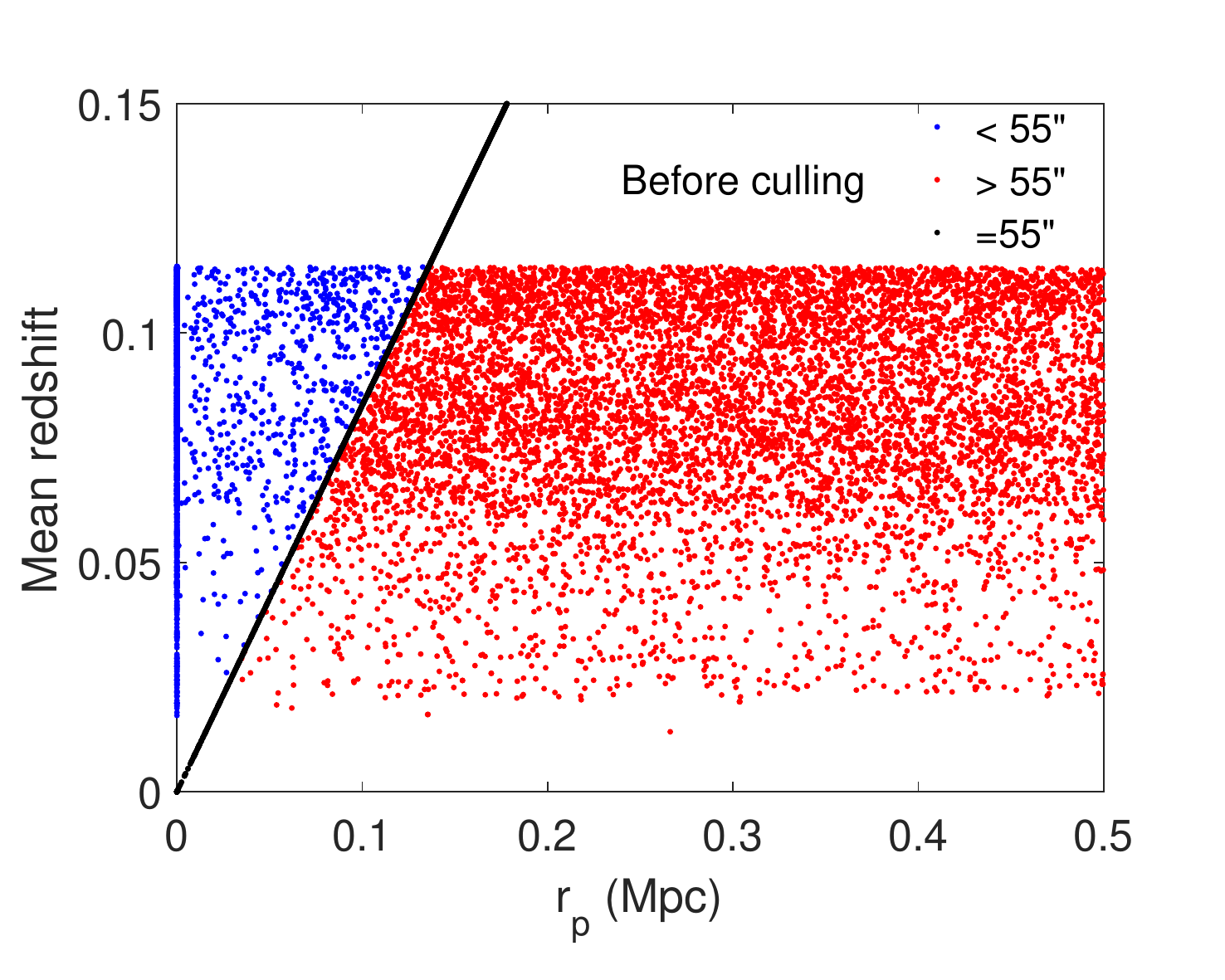}
\includegraphics[width=7cm]{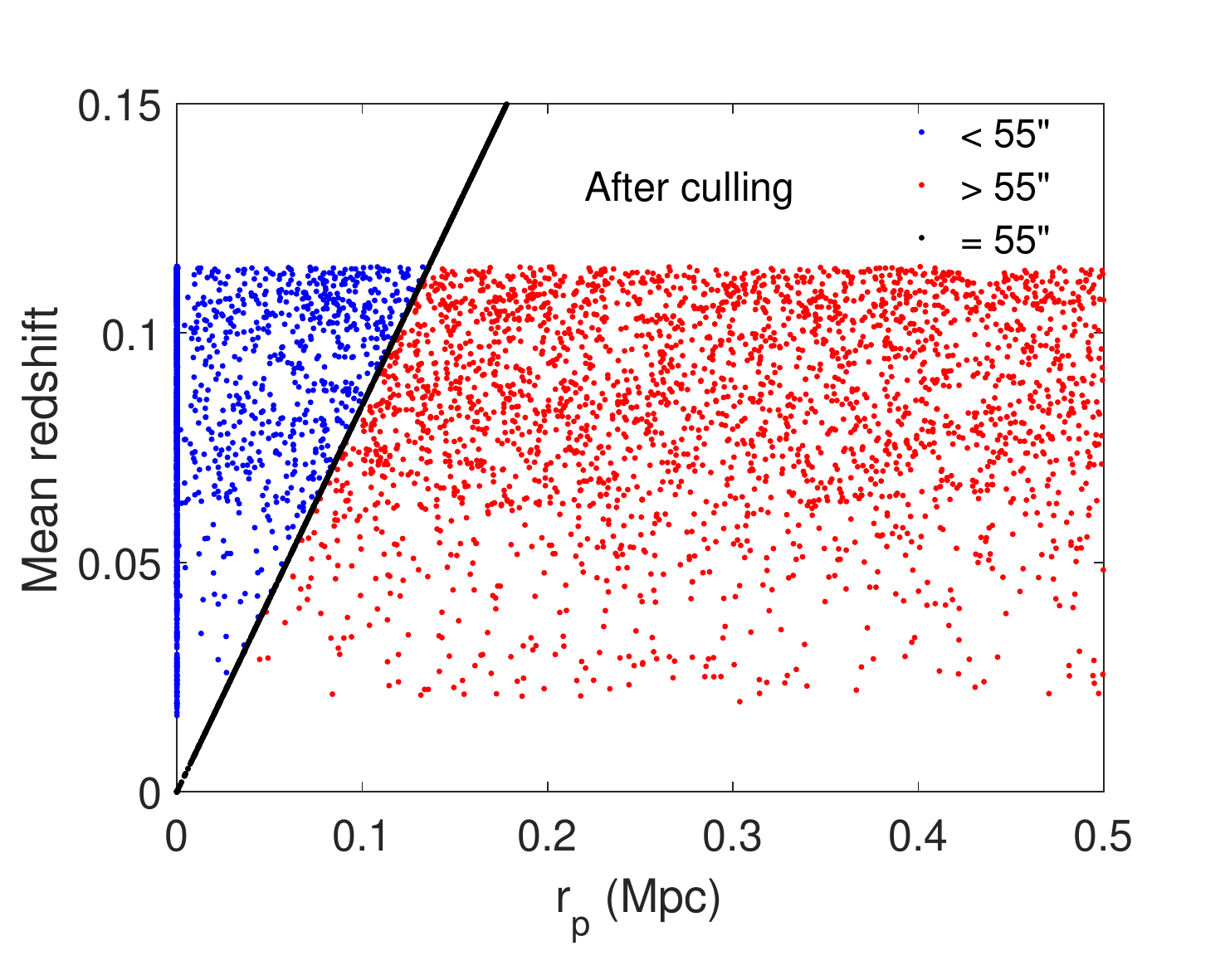}
\caption{The left panel shows the projected separation $r_p$ versus
  mean redshift of galaxy pairs with angular separation
  $<55^{\prime\prime}$ (blue dots) and $>55^{\prime\prime}$ (red dots)
  before culling. The black demarcation line corresponds to an angular
  separation of $55^{\prime\prime}$. The right panel shows the same
  but after the culling procedure.}
\label{fig:cullpair}
\end{figure*}

\begin{figure*}[htbp!]
\centering
\includegraphics[width=7cm]{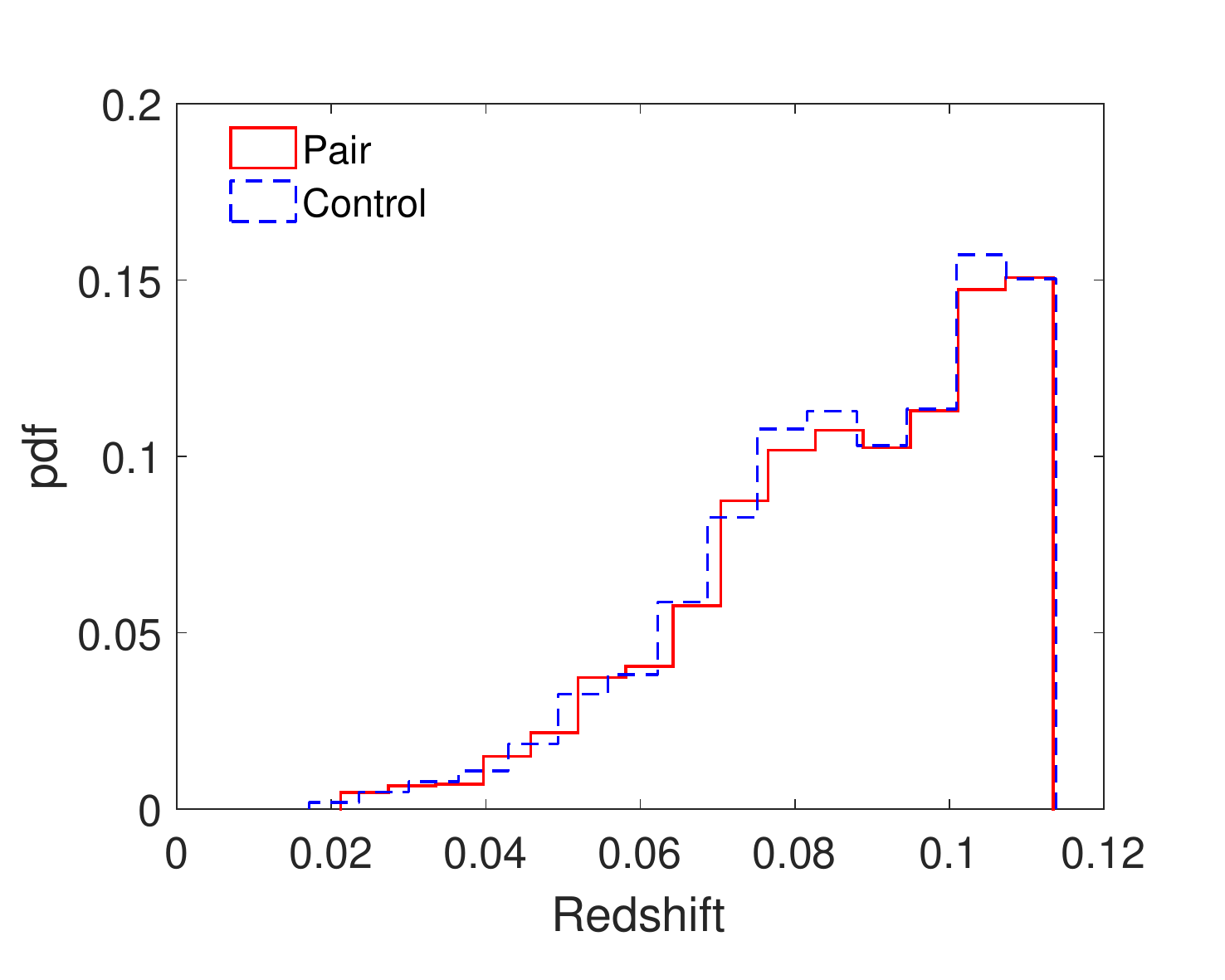}
\vspace{-0.4cm}
\includegraphics[width=7cm]{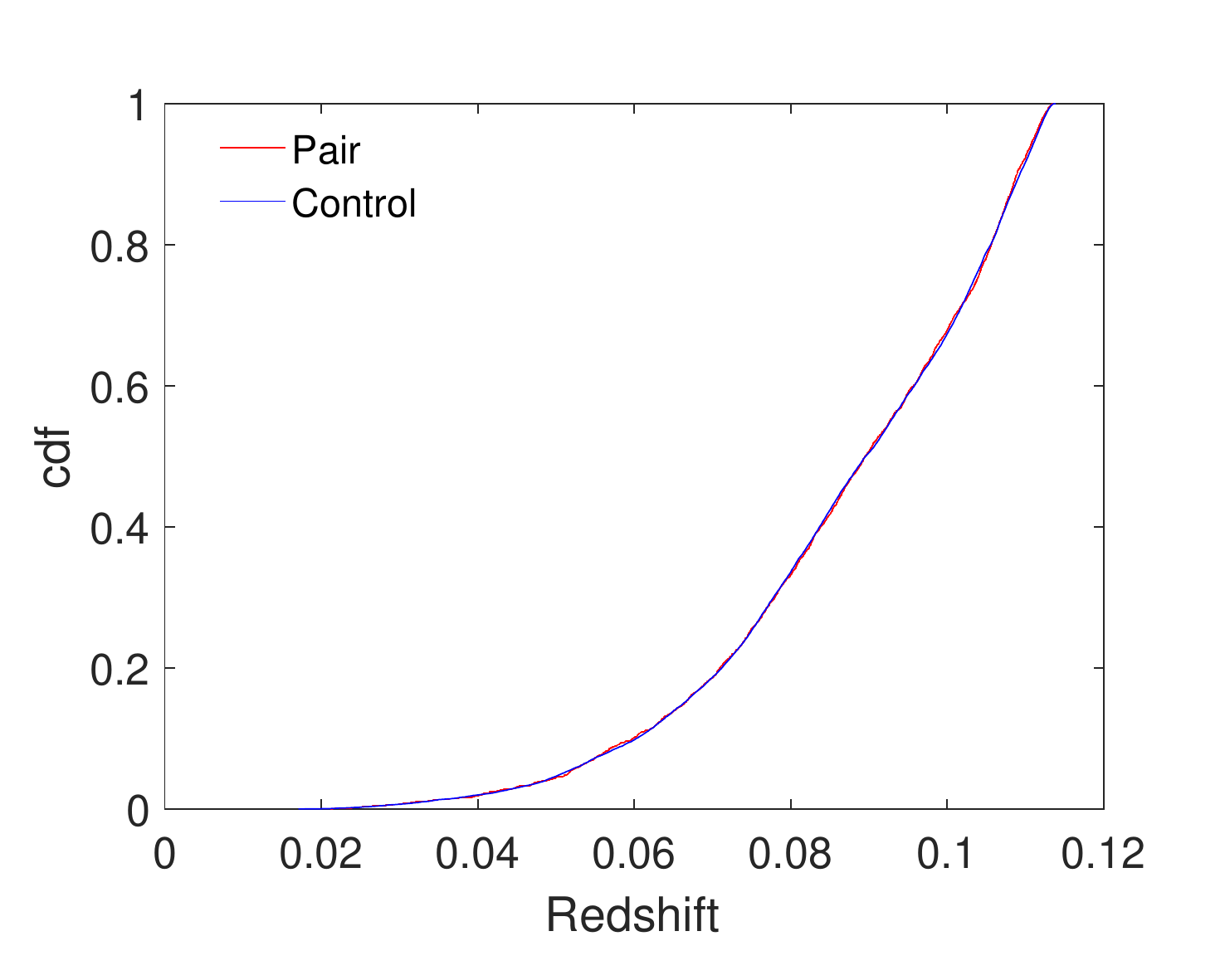}
\includegraphics[width=7cm]{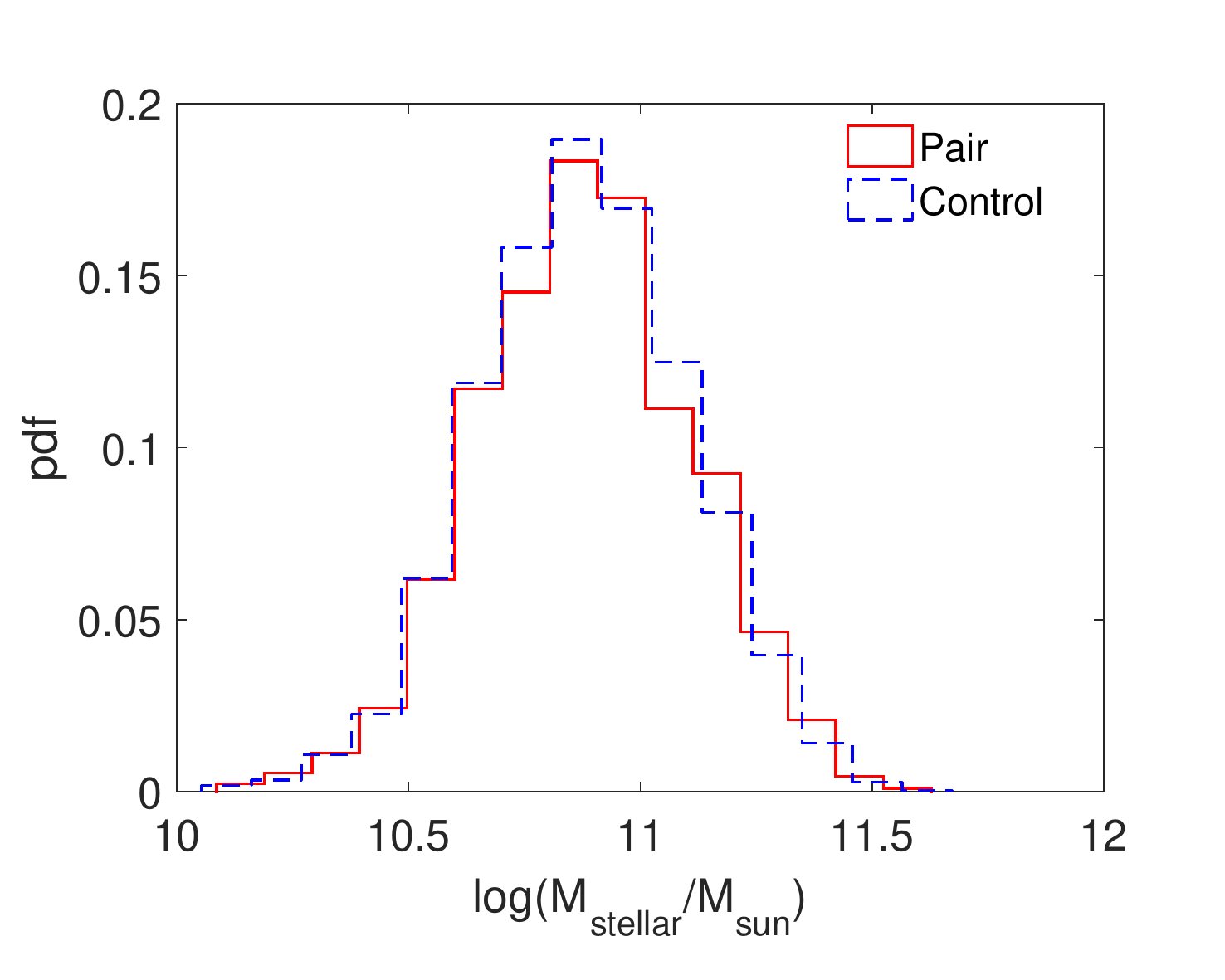}
\vspace{-0.4cm}
\includegraphics[width=7cm]{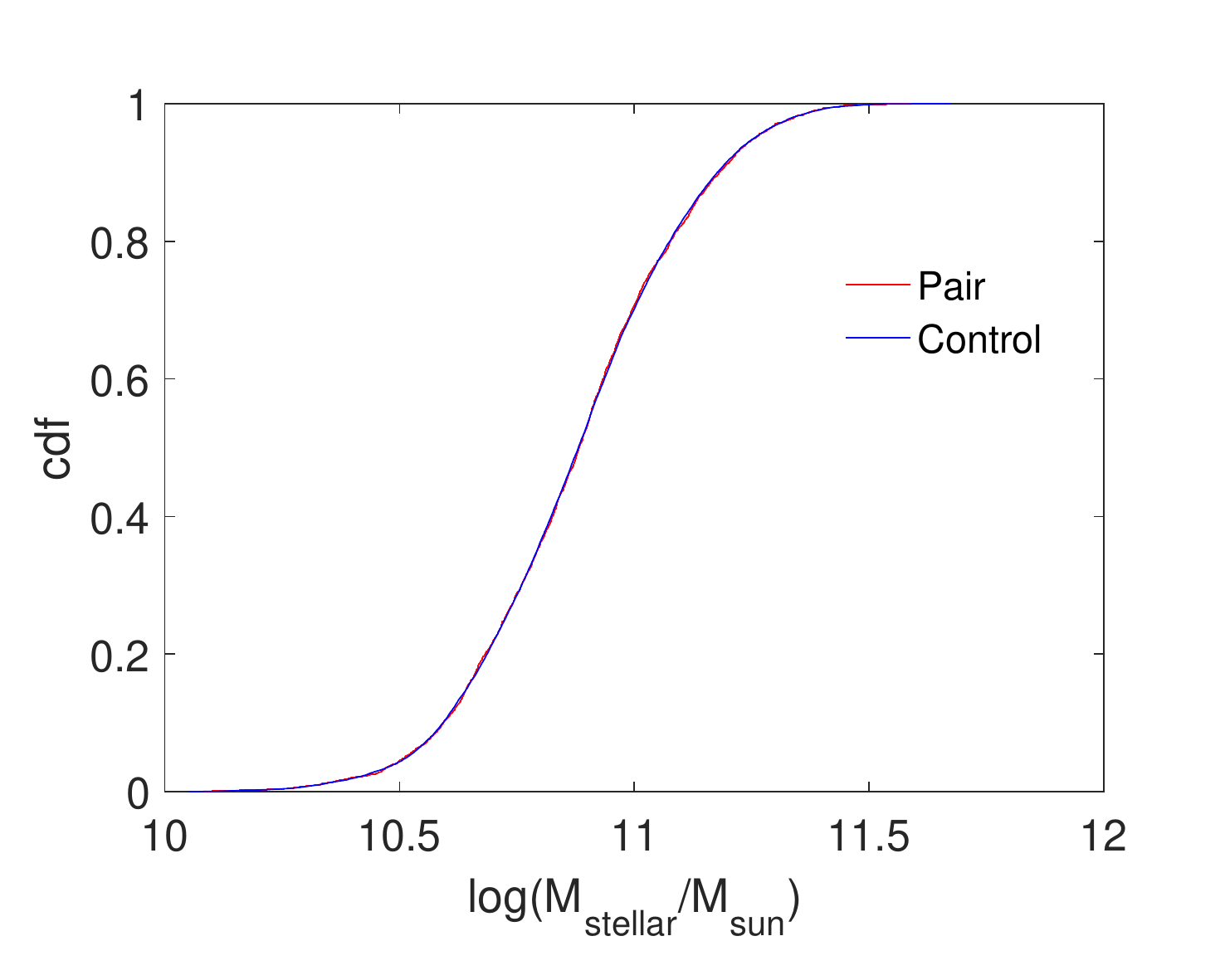}
\includegraphics[width=7cm]{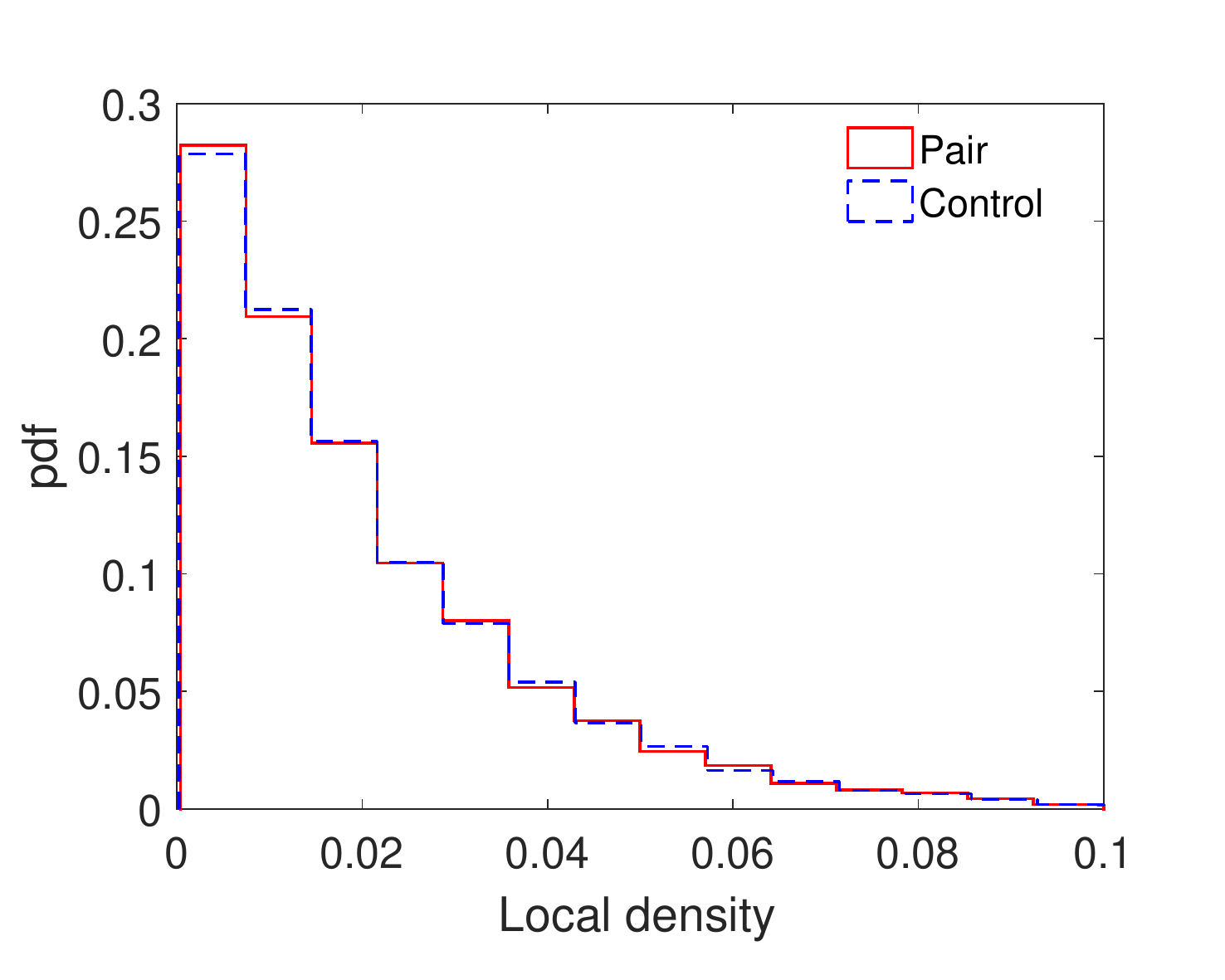}
\includegraphics[width=7cm]{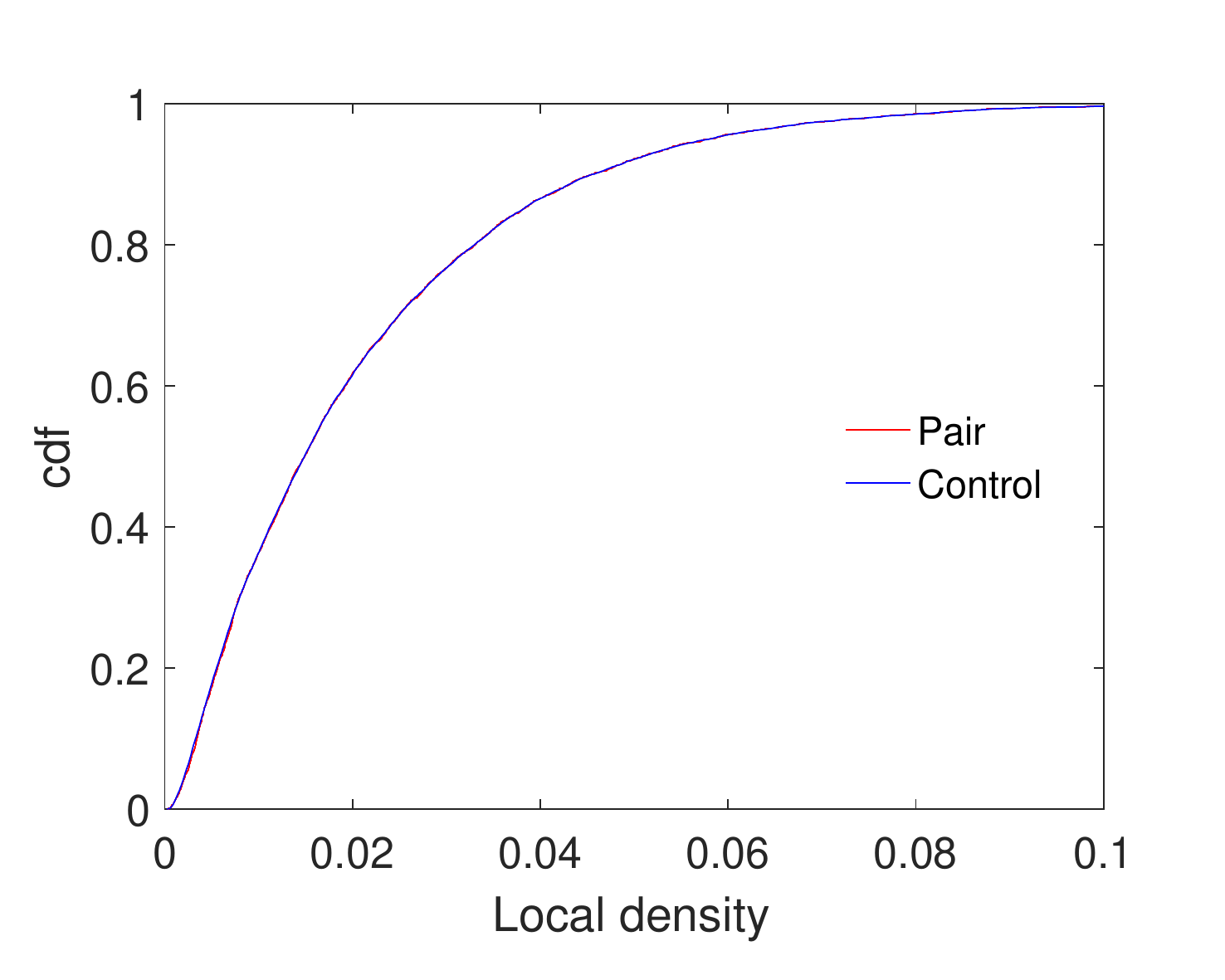}
\caption{The top left, middle left, and bottom left panels,
  respectively show the probability distributions of redshift,
  $log(M_{stellar}/M_{sun})$ and local density for the pair and
  control galaxies. The corresponding cumulative distribution
  functions are compared in the top right, middle right and bottom
  right panels respectively.}
\label{fig:kscontrol}
\end{figure*}

\subsection{Identification of galaxy pairs}

\label{sec:findpairs}

We identify the galaxy pairs using the traditional method based on the
application of simultaneous cuts on the projected separation and the
velocity difference.

We find the projected separation ($r_p$) between any of the two
galaxies in the distribution. We calculate the projected separation
from the redshift of the galaxies using the following relation
\begin{equation}
r_p=\frac{\frac{1}{2}(z_1+z_2) \, \, c}{H_0} \,\,\, \theta
\end{equation}
, where $z_1$ and $z_2$ are the redshifts of the galaxies in pair, $c$
and $H_0$ carry their usual meaning and $\theta$ is their angular
separation given by
                
\begin{equation}                        
\theta =\cos^{-1}\left[ \cos \delta_1 \,\cos \delta_2 \,\cos
  (\alpha_1-\alpha_2)+ \sin \delta_1 \,\sin \delta_2 \right].
\label{theta}
\end{equation}
Here $(\alpha_1, \delta_1)$ and $(\alpha_2, \delta_2)$ are the
equatorial co-ordinates of the two galaxies considered.

The difference between the Hubble velocities of the two galaxies is

\begin{equation}                        
\Delta v = c |z_1-z_2|.
\end{equation} 

In order to select the galaxy pairs, we impose simultaneous cuts on
the projected separation and the velocity difference of the two
galaxies under consideration. Any two galaxies are considered to form
a pair if their projected separation $r_p < 0.5$ Mpc and the velocity
difference $\Delta v<300$ km/s. It is known from earlier studies that
the pairs with $\Delta v>500$ km/s are not likely to be
gravitationally bound \citep{patton00, depropris07} and
interacting. We choose a somewhat larger cut-off for the projected
separation to explore the effects of spurious pairs on larger scales.

An earlier work by Scudder et al. \citep{scudder12b} show that
excluding the galaxies with multiple companions do not alter their
results. So we have allowed a single galaxy to be part of multiple
pairs provided they satisfy the criteria imposed on $r_p$ and $\Delta
v$.

We identify all the galaxy pairs within the same contiguous region and
redshift as the volume limited sample described earlier. The pair
selection algorithm, when applied to this region, yields a total
$114070$ galaxy pairs. We then cross match the galaxies in the volume
limited sample with the galaxies in pairs that provides us with a
total $11287$ galaxy pairs present in our volume limited sample. We
ensure that the matched galaxies in pairs must have measurements of
the stellar mass, star formation rate and internal reddening. We then
impose another cut so as to only consider the pairs with stellar mass
ratio $<10$. This restriction reduces the available number of galaxy
pairs to $11200$.

\subsection{SDSS fibre collision effect: culling pairs}

It is important to take into account the spectroscopic incompleteness
due to the finite size of the SDSS fibres. The minimum separation of
the fibre centres is $55^{\prime\prime}$ due to their finite size
\citep{strauss02}. Consequently, the companion galaxies closer than
$55^{\prime\prime}$ are preferentially missed leading to
under-selection of the close angular pairs. The galaxies within the
collision limit can be still observed if they lie in the overlapping
regions between adjacent plates. The ratio of spectroscopic to
photometric pairs decreases from $\sim 80\%$ at $>55^{\prime\prime}$
to $\sim 26\%$ at lower angular separation \citep{patton08}. This
incompleteness effect can be compensated by randomly culling $67.5\%$
of galaxies in pairs with the angular separation $>55^{\prime\prime}$
\citep{ellison08, patton11, scudder12b}.

We have $11200$ galaxy pairs in our volume limited sample. We find
that there are $8486$ pairs with angular separation (\autoref{theta})
$>55^{\prime\prime}$ and $2714$ pairs with angular separation
$<55^{\prime\prime}$. In a similar spirit to the earlier works, we
randomly exclude $68.02\%$ pairs with $\theta >55^{\prime\prime}$
(\autoref{fig:cullpair}). After the culling procedure, we are left
with a total $5427$ pairs in the volume limited sample.

Our volume limited sample contains $75118$ galaxies that have no
identified pairs according to the criteria applied in
\autoref{sec:findpairs}. We term these galaxies as isolated and use
them to build our control sample as described in the following
subsection.

\begin{table}
\centering
\begin{tabular}{|c |c|c c|}
\hline
& $D_{KS}$ & Confidence level & $D_{KS}(\alpha)$ \\
\hline
Redshift & 0.0120 & 99\% & 0.0288\\
& & 90\% & 0.0216\\
& & 80\% & 0.0190\\
$\log(M_{stellar}/M_{sun})$ & 0.0084 & 70\% & 0.0172\\
& & 60\% & 0.0159\\
& & 50\% & 0.0147\\
& & 40\% & 0.0137\\
Local density & 0.0105 & 30\% & 0.0128\\
\hline
\end{tabular}
\caption{This table shows the Kolmogorov-Smirnov statistic $D_{KS}$
  for comparisons of redshift , $log(M_{stellar}/M_{sun})$ and local
  density of pair and control galaxies. The table also lists the
  critical values $D_{KS}(\alpha)$ above which null hypothesis can be
  rejected at different confidence levels.}
\label{kstab}
\end{table}

\subsection{Building control sample}

The physical properties of interacting galaxies should be compared
only against the control sample of non-interacting galaxies to detect
the impact of tidal interactions. The colour and SFR of galaxies
depends on their stellar mass and environment. So it is crucial to
ensure that the distributions of stellar mass and local density for
the pairs and control samples are statistically indistinguishable. The
colour and the star formation activity of galaxies are also known to
depend on the redshift. Besides, the redshift dependent selection
effects can not be eliminated completely even in a volume limited
sample. So we also decide to match the redshift distributions of the
paired galaxies and control sample of isolated galaxies.

For estimating the local density, we find the distance to the $k^{th}$
nearest neighbour from a galaxy. The $k^{th}$ nearest neighbour
density \citep{casertano85} around a galaxy is defined as
\begin{equation}
\eta_k = \frac{k-1}{V(r_k)}
\end{equation}
Here $r_k$ is the distance to the $k^{th}$ nearest neighbour and
$V(r_k) = \frac{4}{3}\pi r^3_k$ is the volume of the sphere having a
radius $r_k$. We have considered the distance to the $5^{th}$ nearest
neighbour from each galaxy by taking $k=5$ in this work. The local
density of galaxies can be underestimated near the survey boundary.
We consider only those galaxies for which $r_k < r_b$. Here $r_b$ is
the closest distance of the galaxy from the survey boundary. We
calculate the local density for all the galaxies in our volume limited
sample for which the criterion mentioned earlier is satisfied. We
finally get the local density estimates for $8886$ paired galaxies and
$69308$ isolated galaxies in our volume limited sample.

We have $5427$ galaxy pairs in our volume limited sample after
correcting for the fibre collision effect. We have adopted a strategy
similar to Ellison et al. 2008 \citep{ellison08} for building the
control sample. We build the control sample of the isolated galaxies
by simultaneously matching their stellar mass, local density and
redshift. For each paired galaxy, we pick $5$ unique isolated galaxies
matched in stellar mass, local density and redshift. We match the
paired galaxies and their controls to within 0.08 dex in stellar mass,
$0.001$ in local density and $0.005$ in redshift. We also
simultaneously ensure that only the isolated galaxies that do not have
a companion within $\leq 1$ Mpc can be part of the control sample of
non-interacting galaxies. We match every paired galaxy to their
controls and then perform a Kolmogorov-Smirnov (KS) test on the
stellar mass, local density and redshift distributions
(\autoref{fig:kscontrol}). The control samples are accepted only when
their stellar mass, local density and redshift distributions are
consistent with that for the paired galaxies at a level of at least
$30\%$ KS probability (\autoref{kstab}). This implies that the null
hypothesis can not be rejected at $\geq 30\%$ confidence level, and
the galaxy pair and control samples are highly likely to be drawn from
the same parent distribution. The control matching in stellar mass,
local density and redshift eliminate most of the biases that can
plague a comparison between the two samples \citep{perez09b}. Any
other systematic biases equally affect both the samples and should not
be a matter of concern here.

Only the paired galaxies with measurements of stellar mass, local
density and redshift can be used for the KS test required in our
analysis. Further, the condition that there should be at least $5$
control galaxies for each paired galaxy reduces the number of
available galaxy pairs.  Finally, we have $2011$ galaxy pairs that
fulfil all these criteria. These $2011$ galaxy pairs are formed by a
total $3835$ galaxies. The control sample of the paired galaxies
consists of a total $19175$ isolated galaxies.

We define the major pairs that have the stellar mass ratio in the
range $1 \leq \frac{M_1}{M_2} < 3$. All the pairs with stellar mass
ratio $3 \leq \frac{M_1}{M_2} \leq 10$ are identified as minor
pairs. We classify the $2011$ galaxy pairs according to this criteria
and find that there are $1831$ major pairs and $180$ minor pairs. We
further note that $649$ major pairs have a projected separation
$r_p=0$. This can happen if the two members in a pair have a non-zero
radial separation but lie along the same line of sight.  Such cases
may also arise in a merger system where the angular separation between
the two members are extremely small. Unreliable deblending can also
lead to false identification of single galaxies as close pairs. It is
difficult to ascertain the $r_p=0$ pairs as mergers based on pair
separation alone. So we discard these pairs from our analysis. We also
drop the corresponding control galaxies from our control sample. We
finally have $1182$ major pairs and $180$ minor pairs in our volume
limited sample, that have measurements of dust corrected colour, SFR
and local density.

\begin{figure*}[htbp!]
\centering
%\includegraphics[width=7cm]{sfr_rp_all_maj_min.pdf}
%\vspace{-0.4cm}
%\includegraphics[width=7cm]{clr_rp_all_maj_min.pdf}
\includegraphics[width=7cm]{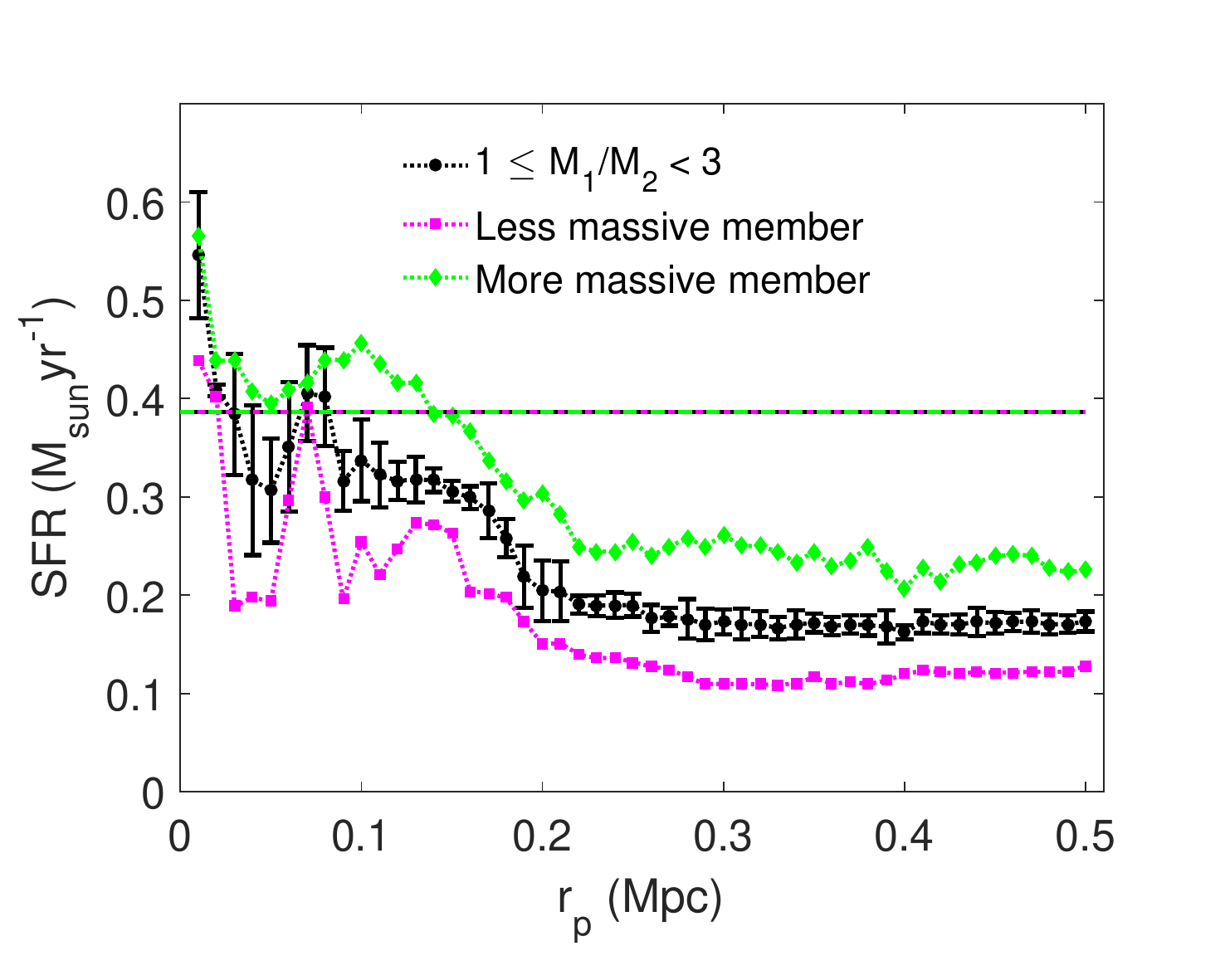}
\vspace{-0.4cm}
\includegraphics[width=7cm]{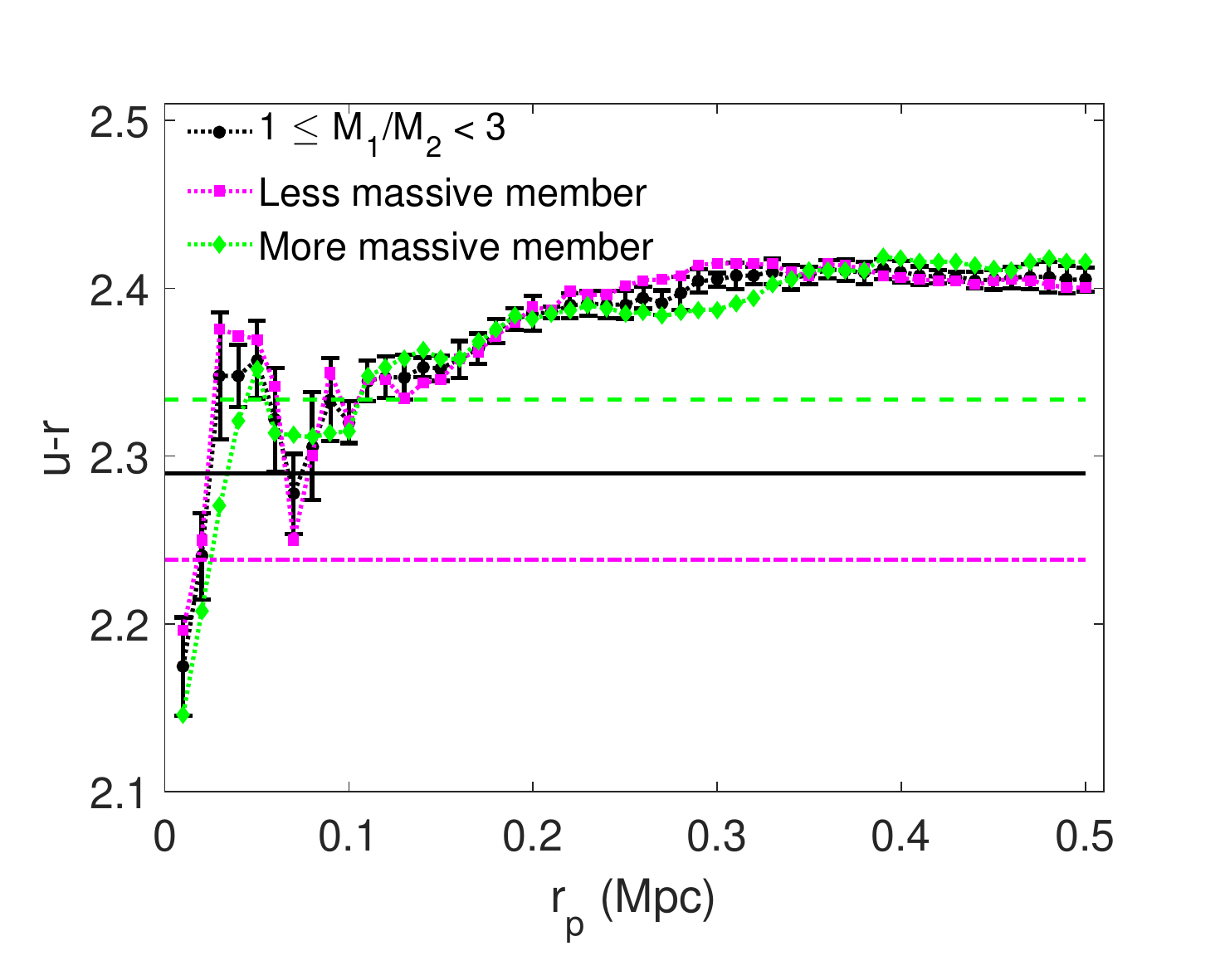}
\includegraphics[width=7cm]{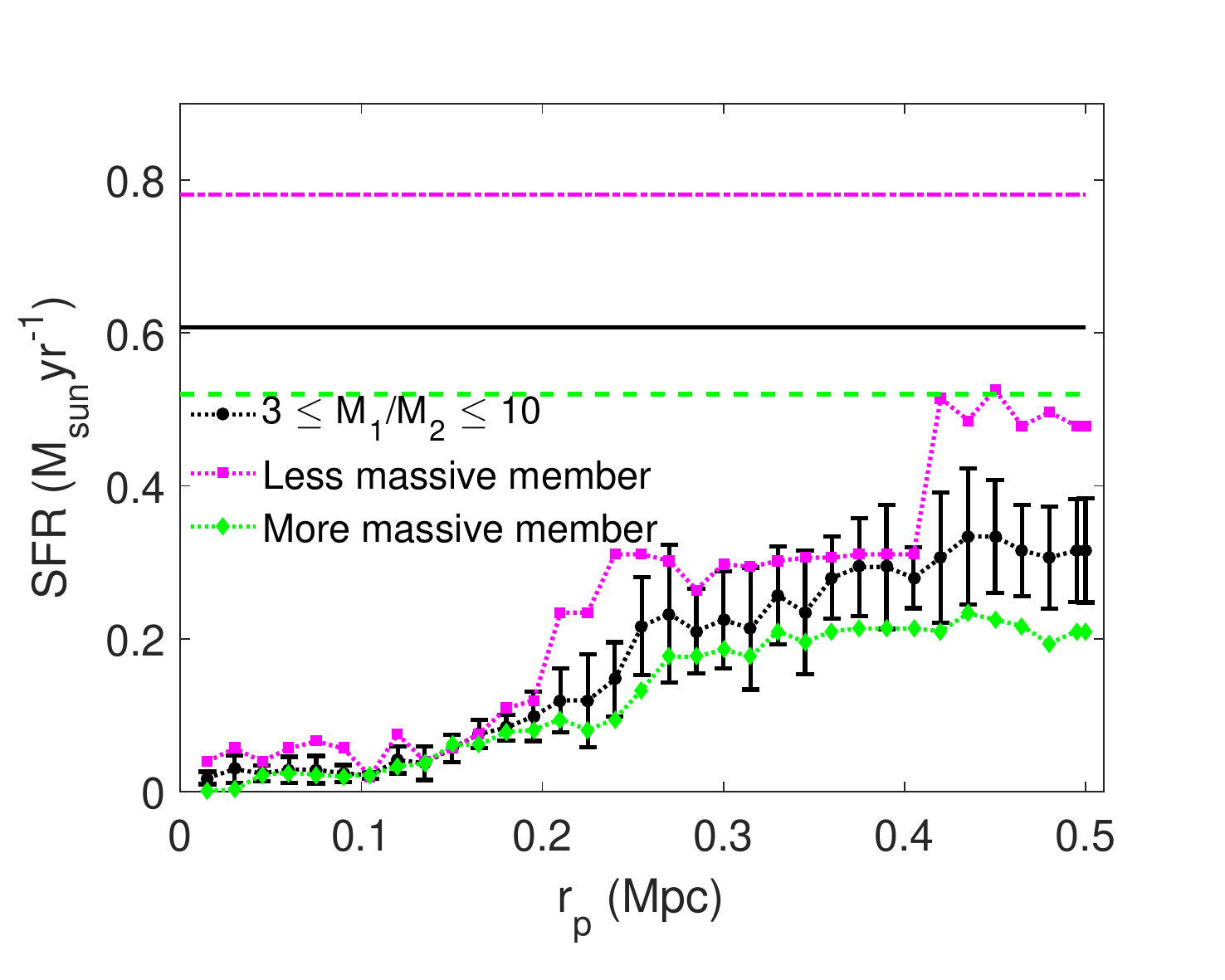}
\includegraphics[width=7cm]{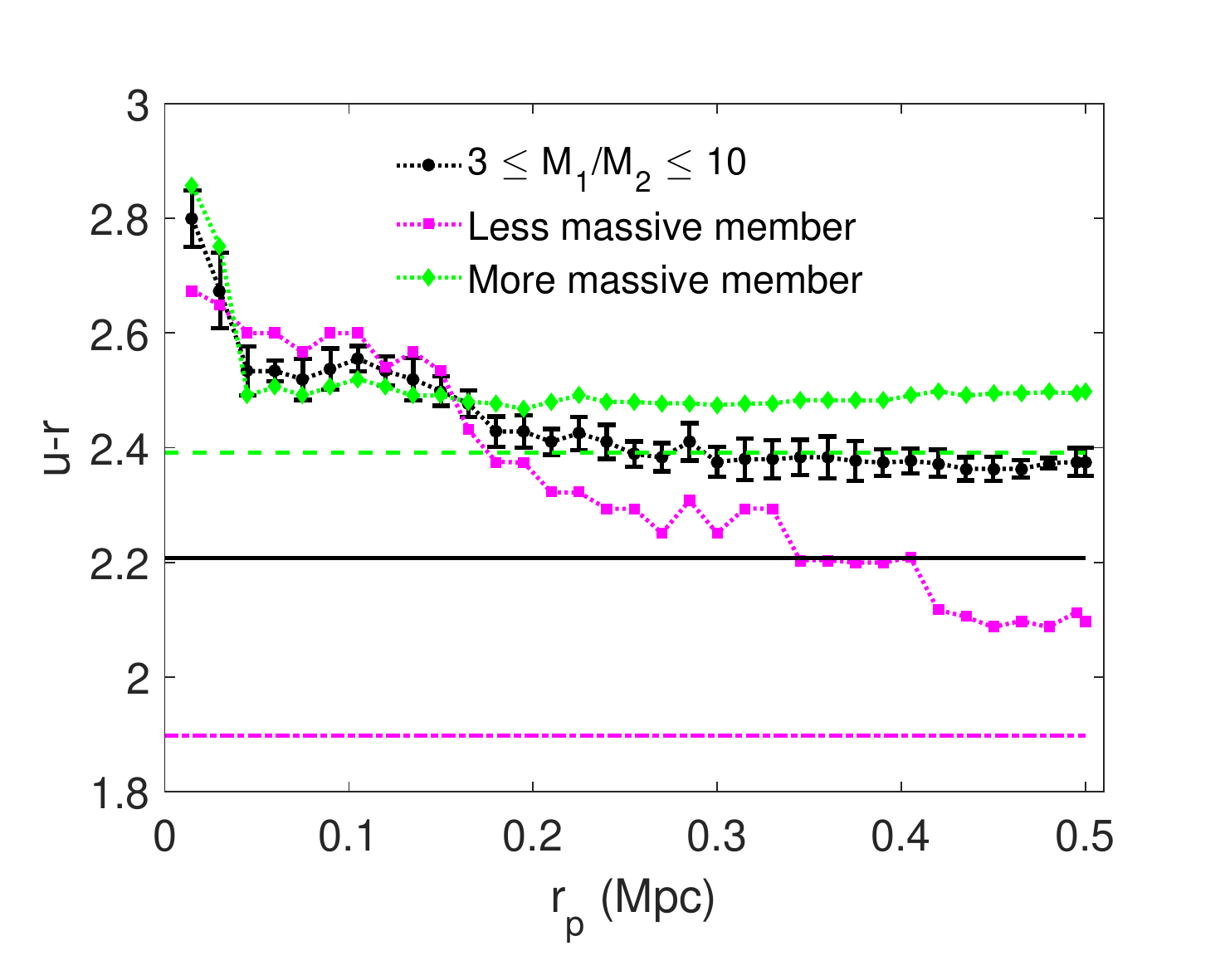}
\caption{The top left panel shows the cumulative median SFR for major
  pairs as a function of the projected separation. The results for all
  the major pair is shown using the black curve. The $1-\sigma$ error
  bars shown here are obtained using $10$ jackknife samples from the
  original data. The results for the more massive and less massive
  members in major pairs are shown separately in the same plot using
  the green and pink curves, respectively. The $1-\sigma$ error bars
  for the more massive and less massive members in major pairs are
  comparable to that shown for all members in major pairs (black
  curve). We do not show these errorbars here for the sake of
  clarity. The median SFR for the control samples of each set of
  galaxies are shown with the horizontal straight lines. The bottom
  left panel shows the cumulative median SFR as a function of
  projected separation for the minor pairs. We show the cumulative
  median (u-r) colour as a function of projected separation for the
  major and minor pairs in the top right and bottom right panels. The
  results for the more massive and less massive members of the galaxy
  pairs are shown together in each of these panels.}
\label{fig:majmin}
\end{figure*}

\begin{figure*}[htbp!]
\centering
%\includegraphics[width=7cm]{sfr_rp_all_lden.pdf}
%\vspace{-0.4cm}
%\includegraphics[width=7cm]{clr_rp_all_lden.pdf}
\includegraphics[width=7cm]{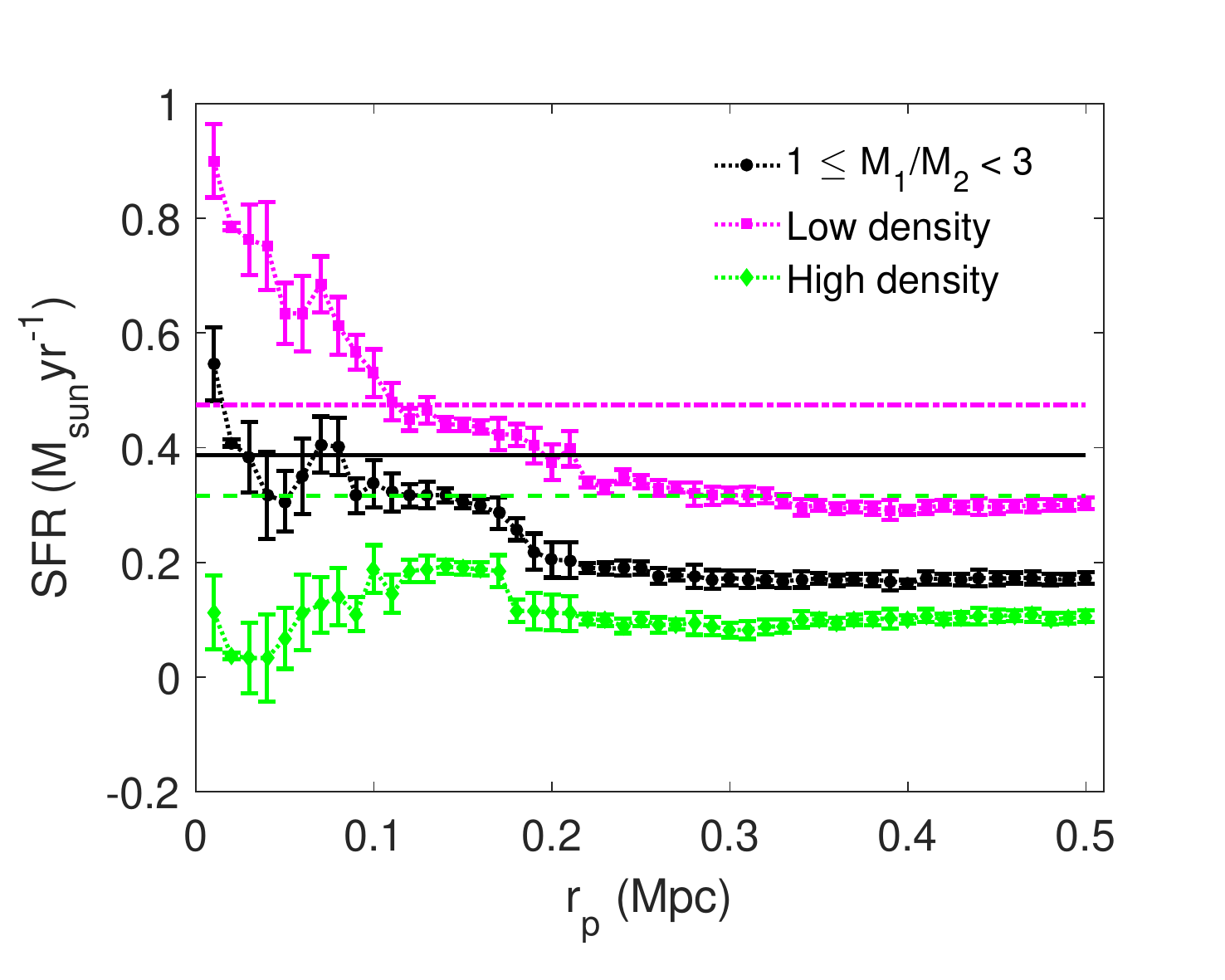}
\vspace{-0.4cm}
\includegraphics[width=7cm]{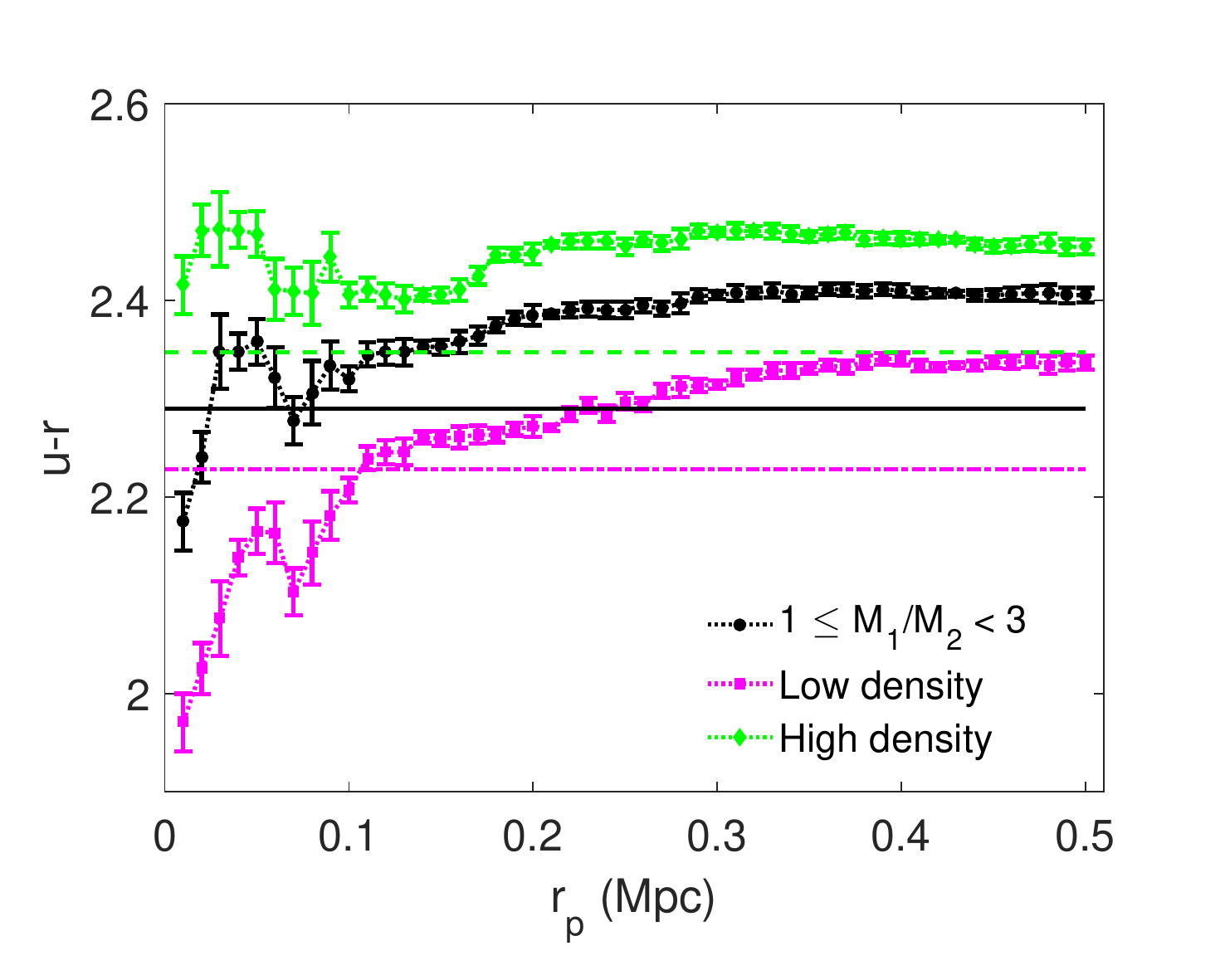}
\includegraphics[width=7cm]{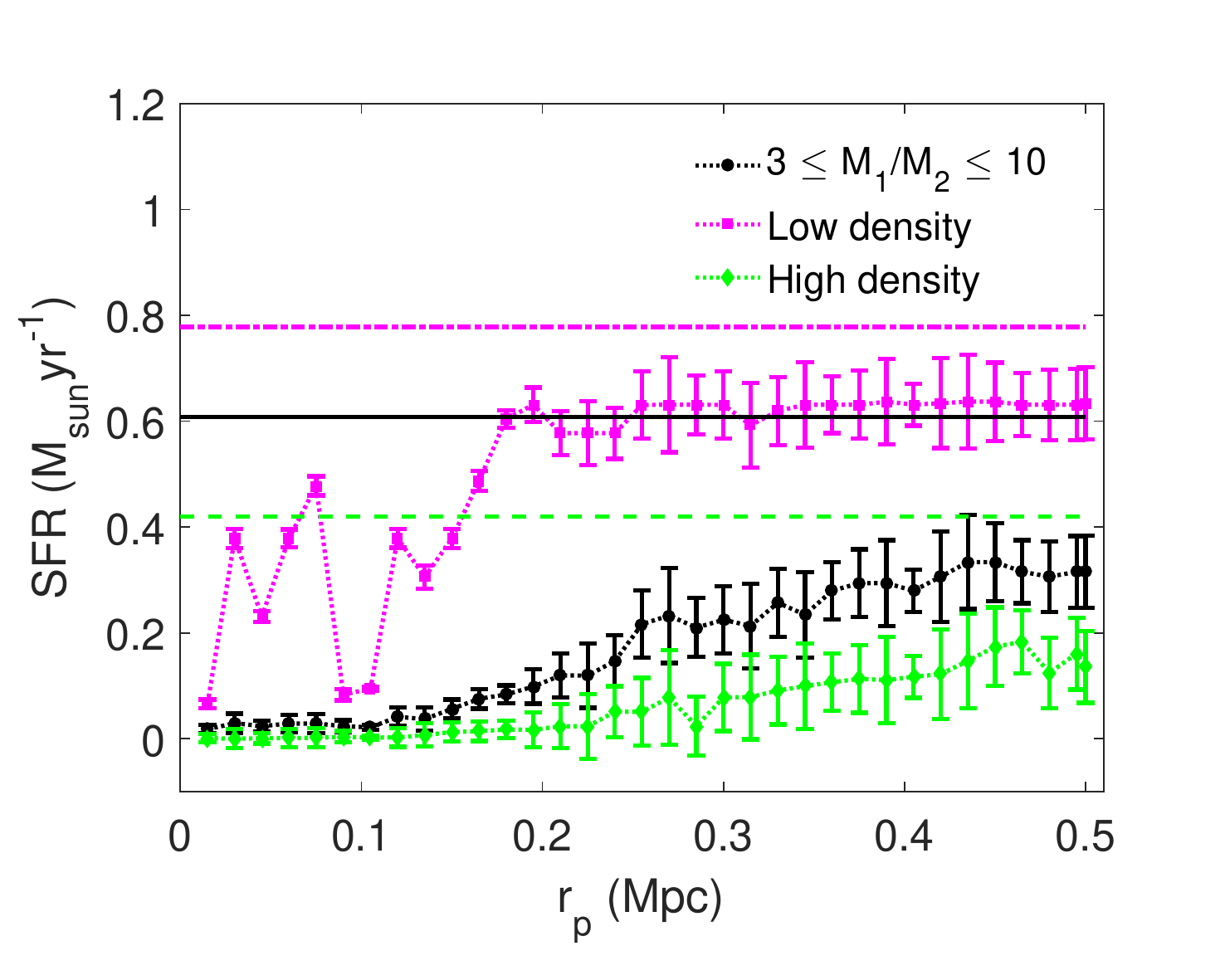}
\includegraphics[width=7cm]{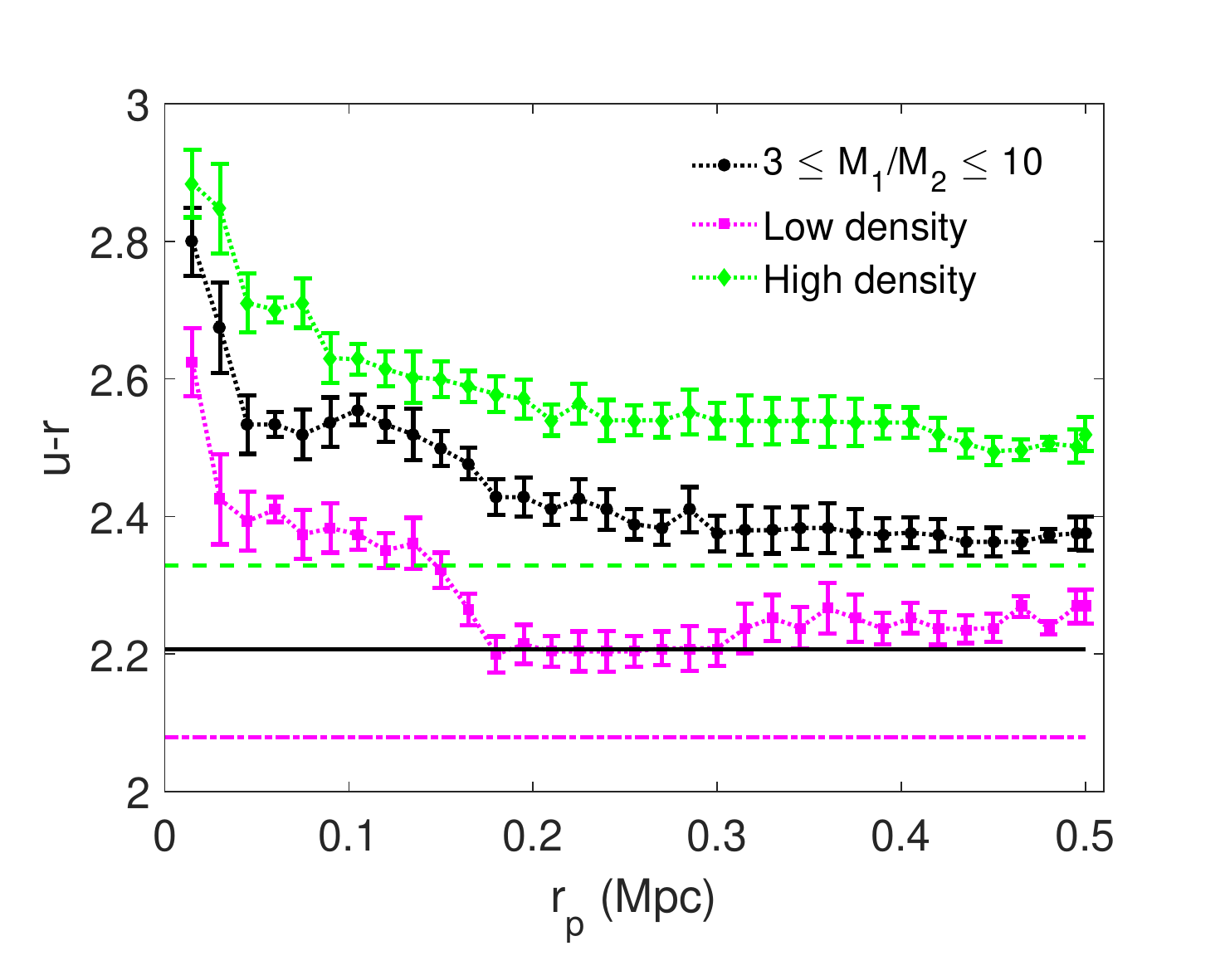}
\caption{Same as \autoref{fig:majmin} but here we show together with
  the results for the major and minor pairs at high-density and
  low-density environments.}
\label{fig:envpair}
\end{figure*}

\section{Results}

\subsection{Impact of major and minor interactions on star formation rate and colour}

We show the cumulative median of the SFR and the dust corrected
$(u-r)$ colour of the major pairs as a function of the projected
separation respectively in the top left and top right panels of
\autoref{fig:majmin}. The black curves in these two panels show the
results for all the major pairs in our volume limited sample. The
$1-\sigma$ error bars are obtained using $10$ jackknife samples
prepared from the original data. The horizontal black lines in the top
two panels show the respective medians for their control samples. In
the top left panel, we find that the cumulative median SFR of the
major pairs increases with decreasing pair separation and rises above
the control median below $30$ kpc. It is interesting to note that the
same trend is also observed in the top right panel, where the dust
corrected $(u-r)$ colour of the major pairs decreases with decreasing
separation and goes below the control median below $30$ kpc. An
increase in the SFR corresponds to a decrease in the $(u-r)$
colour. So the top two panels together indicate that the tidal
interactions in major pairs trigger starburst at closer pair
separation. The enhancement is most pronounced at projected separation
$<30$ kpc. 

It is also interesting to note the presence of a bump in both the top
panels. We find that both the cumulative SFR and $(u-r)$ colour show
an upturn and a downturn at $70$ kpc where they intersect their
control medians. The fact that these curves intersect their control
medians at $\sim 30$ kpc and $70$ kpc suggest the presence of a
transition region where the tidal interaction can impact the SFR and
colour in both directions. The presence of identical features for both
the SFR and colour suggest that this can not arise due to statistical
fluctuations or chance appearance. We note that such an upturn and
downturn in the SFR at $50$ kpc have been already reported by several
earlier studies \citep{lambas03, nikolic04, ellison08}. The number of
non-interacting false pairs increase at larger separation due to the
projection effects. The presence of the transition region marked by
the upturn and downturn is generally ascribed to the contaminations
from the projection effects.

The effect of the tidal interaction diminishes with the increasing
pair separation. A gradual change can be seen up to $200$ kpc in both
the top panels. Both the SFR and the $(u-r)$ colour plateaus out to
nearly constant values beyond a separation of $200$ kpc. However, as
noted earlier, the pairs with wider separation are likely to be
contaminated from the projection effects. These pairs may not truly
represent either the field population or the truly interacting
systems. It can be clearly seen from the deviation of the cumulative
medians from the respective control medians even at wider separation.

We also show together with the results for the more massive and less
massive members in major pairs in the top panels of
\autoref{fig:majmin}. Our results indicate that the tidal interactions
impart a greater influence on the more massive members in major pairs.
We note that the median SFRs in the control samples of the more
massive and less massive members in major pairs are nearly identical.
This can be attributed to the fact that the heavier and lighter
members in major pairs have similar stellar masses that are within a
factor of $<3$. It also implies that such variations in mass do not
cause a noticeable differences in the SFR of the isolated galaxies.

The effects of tidal interactions on the minor pairs in our volume
limited sample are shown in the two bottom panels of
\autoref{fig:majmin}. In the bottom left panel of this figure, we show
the cumulative median SFR of the minor pairs as a function of the
projected separation. We find that the SFR show a gradual decrease
with the decreasing separation, and the SFR in minor pairs are lower
than their control median at all separations. These indicate that the
tidal interactions in minor pairs suppress the star formation in the
intrinsically brighter galaxy pairs. This suppression is contrary to
the tidally triggered SFR enhancement in major pairs. The median SFRs
in the control samples of the more massive and less massive members in
minor pairs are markedly different. This is caused by larger mass
differences between the two members in minor pairs. The bottom right
panel showing the cumulative $(u-r)$ colour as a function of the pair
separation also exhibits the same trend as SFR. We see that the minor
pairs become redder with decreasing pair separation. We note that both
the more massive and less massive members in minor pairs suppress
their star formation and become redder at closer separation. In both
the bottom panels of \autoref{fig:majmin}, we note that the magnitudes
of the deviations from the corresponding control medians are larger
for the less massive members in minor pairs. This clearly indicates
that the minor interactions influence the SFR and colour of the
lighter members in pairs to a greater extent. The minor interactions
are usually known to trigger star formation in the low luminosity star
forming galaxies. We consider only the brighter galaxies with a
relatively lower SFR and redder colour in our sample that can be
clearly seen from the median SFR and colour of the major and minor
pairs and their controls in different panels of
\autoref{fig:majmin}. Our results indicate that the minor interactions
quench star formation in the intrinsically brighter galaxy pairs.

\subsection{Environmental dependence of major and minor interactions}

We address the environmental dependence of the major and minor
interactions in \autoref{fig:envpair}. We use the local density at the
location of each paired galaxy to characterize their environment. Once
the pairs are classified as major and minor, we sort the local
densities for each type of galaxy pair and determine the respective
median densities. We use the median density as a cut that divides each
sample into two subsamples, each comprising an equal number of
galaxies. We define the densities above and below the median as high
and low, respectively.

Besides the results for the entire sample of major and minor pairs, we
also show together with the results for these pairs in high-density
and low-density environments in different panels of
\autoref{fig:envpair}. The black curves in each of the four panels of
\autoref{fig:envpair} are the same as those shown in the corresponding
panels of \autoref{fig:majmin}. In each panel, the results for the
paired galaxies in low-density and high-density environments are
separately shown using pink and green colours. The top two and the
bottom two panels of \autoref{fig:envpair} show the results for the
major pairs and minor pairs, respectively. The top left panel of
\autoref{fig:envpair} show that the cumulative median SFR for major
pairs in the low-density regions is significantly higher than their
control median for smaller pair separations. It decreases with the
increasing separation but remains above the corresponding control
median up to at least $100$ kpc. On the other hand, the SFR of the
major pairs hosted in the high density regions remains below their
control median throughout the entire length scale.  This result
suggests that the SFR can get suppressed in the major pairs in the
high-density regions. The top right panel of \autoref{fig:envpair}
show the results for the cumulative median $(u-r)$ colour. We find the
same trend with the galaxy colour of major pairs. The major pairs in
the low-density regions become bluer with decreasing pair separation,
and the colours for these pairs stay below their control median up to
$100$ kpc. The colour of the major pairs in high-density regions
remains above their control median for the entire length scale that
indicates that the major pairs in the high-density regions are redder
than their isolated controls in similar environments. The results in
the top two panels of \autoref{fig:envpair} thus tell us that the
major interactions can both trigger starburst or quench star formation
depending on their environment. It may be noted that the median SFRs
in the control samples of major pairs at the low density and high
density regions are different due to the environmental dependence of
star formation.

We show the results for the minor pairs in the high-density and
low-density environments in the bottom two panels of
\autoref{fig:envpair}. The bottom left panel shows the cumulative
median SFR, and the bottom right panel shows the cumulative median
$(u-r)$ colour as a function of the projected separation. We find that
all the curves in the bottom left panel stay below their control
median, and all the curves in the bottom right panel lie above their
control median for the entire range of pair separation. The SFR
progressively diminishes, and the colour gradually reddens with the
decreasing pair separation at low and high-density regions. These
results indicate that the minor interactions in brighter pairs always
quench their star formation irrespective of their environments.  The
bottom two panels also clearly show that quenching in minor pairs is
more pronounced in the high-density regions.

\section{Conclusions}

We study the effects of tidal interactions on star formation rate, and
the dust corrected $(u-r)$ colour of paired galaxies using a volume
limited sample from the SDSS. We classify the galaxy pairs as `major'
and `minor' based on their stellar mass ratio and then study the SFR
and colour in these pairs as a function of the projected separation.
These results are then compared against those obtained from the
control samples of isolated galaxies. We prepare the control samples
by matching the redshift, stellar mass and local density of the paired
galaxies in our sample. The tidal interactions may not have an equal
impact on both the members of a galaxy pair. So we study the effects
of the tidal interaction on the properties of the more massive and
less massive members in a pair. Further, the environment is known to
be an important driver of galaxy evolution. We separately study the
impact of major and minor interactions on SFR and colour in high
density and low-density environments. The main conclusions of this
analysis are as follows:

(i) We find that the major interactions significantly enhance the SFR
in paired galaxies up to at least $30$ kpc. The efficiency of the
tidally triggered star formation activity diminishes with the
increasing pair separation. The tidal interaction seems to influence
SFR even at larger length scales. However, this is difficult to
confirm due to significant contamination from spurious pairs at wider
separations. Besides the SFR, the dust corrected $(u-r)$ colour is
strongly affected by the major interactions. The major galaxy pairs
become increasingly bluer at smaller projected separations. The
interaction induced change in the colour of the major pairs also
extends up to $30$ kpc. The impact of major interactions on colour
shows the same trend as those observed for the SFR. We find that the
more massive members in major pairs exhibit a more significant change
in SFR and colour.

(ii) The minor interactions in galaxies with higher intrinsic
luminosity suppress the SFR in both members in a pair. The suppression
of the SFR in minor pairs increases with the decreasing pair
separation. The changes are most prominent within $100$ kpc but are
present throughout the entire length scale probed. The results at
larger pair separations are contaminated by the projection effects and
are less reliable. The colour of the minor pairs becomes increasingly
redder at smaller pair separations. These results together suggest
that the minor interactions quench the star formation in intrinsically
brighter galaxy pairs. Interestingly, we find that the impact of tidal
interactions are more significant in the less massive members of the
minor pairs.

(iii) The environment characterized by the local density plays a
crucial role in influencing the star formation activity and colour of
the galaxy pairs. We find that SFR enhancement due to major
interaction becomes more pronounced in the low-density regions. The
sample of major pairs in low-density exhibits a marked increase in
star formation activity than the undivided sample of major pairs. The
enhancement of star formation activity in such pairs extends to a
larger length scale up to $100$ kpc. Contrary to this, the major pairs
in the high-density regions significantly decrease star formation
activity at all pair separations. We observe a similar trend with the
colour of major pairs in low-density and high-density regions. The
major pairs in the low-density environment become increasingly bluer
at smaller pair separations when compared with their control
samples. The changes in colour in the low-density sample are much more
pronounced than its parent sample and extend out to $100$ kpc. The
major pairs in the high-density regions are significantly redder than
their control galaxies at all pair separations. These results indicate
that the major interaction between galaxies can enhance or quench star
formation depending on their embedding environments.

(iv) The minor interactions between intrinsically brighter galaxies
always suppress the star formation irrespective of their environment.
We find that the minor pairs at both the low-density and high-density
regions are significantly less star forming and redder as compared to
their control matched counterparts.

We now compare our results with the earlier studies and discuss the
implications of our findings. We observe that a significant
enhancement in SFR of paired galaxies within the projected separation
of $30$ kpc is consistent with most of the previous studies. Several
earlier works studied the role of environment on interaction induced
SFR enhancement in galaxy pairs. Lambas et al. \citep{lambas03}
studied the effects of galaxy interactions in the field using 2dFGRS
and find that the SFR for paired galaxies is significantly enhanced
compared to the isolated galaxies. They also noted that these
enhancements are more significant for the brighter galaxies in a
pair. Some other studies \citep{ellison08, ellison10} analyzed the
SDSS galaxies in the low density regions to reach a similar
conclusion. There are observational evidences that SFR in paired
galaxies may be influenced by tidal interactions up to $150$ kpc
\citep{scudder12b, patton13}. Our results for the major interactions
in the low-density environments are consistent with these results. The
higher fraction of gas-rich and bulge-less galaxies in the low-density
environments favour the greater efficiency of tidally triggered star
formation. It has been found that the close passage between bulge-less
galaxies are the most efficient in triggering star formation
\citep{cox09}. Alonso et al. \citep{alonso04} studied the effects of
galaxy interactions in groups and clusters using the 2dFGRS data and
reported that the interaction induced star formation enhancement is
less efficient in the high-density regions. Their results show that
the galaxy pairs in rich groups are systematically redder than other
group members. The dominance of the gas-poor and bulge dominated
systems in the high density environments may be the primary reason for
the lower efficiency of interaction induced SFR enhancement. It is
known that the interaction induced star formation are less efficient
in bulge-dominated galaxies as the central bulge provides stability
against tidal torques \citep{binney87}.

Interestingly, we observe a quenching of star formation in major pairs
at high-density environments. This result is different from the
earlier observations and may have several implications. The galaxy
interactions in denser environments may have been very efficient in
the past, leaving very little gas for new star formation. The galaxies
at early stage of their evolution lack stability and consequently the
effectiveness of the tidal interaction in inducing star formation is
larger during this period \citep{tissera02}. Further, the tidal
interactions may also cause gas loss through AGN or shock driven winds
\citep{murray05, springel05}, induce bar quenching \citep{haywood16}
and morphological quenching \citep{martig09}. It may also be noted
that we consider only the intrinsically brighter galaxies in our
analysis. The quenching in these pairs in the high density
environments indicates that the intrinsic luminosity of the galaxy
pairs are also important besides their luminosity or mass ratio. Our
results imply that the major interactions can enhance or suppress the
star formation in galaxy pairs based on their environments and
luminosity. Such interaction induced effects in both directions can
contribute to the observed bimodality in colour and other galaxy
properties.

The minor interactions are known to trigger a mild enhancement in the
star formation activity. Ellison et al. \citep{ellison08} noted that
the less luminous galaxies in minor pairs are more susceptible to
enhanced star formation. We find that the minor interactions in the
intrinsically brighter galaxy pairs quench star formation irrespective
of their environment. The more luminous galaxies in the intrinsically
brighter pairs have higher stellar mass and are generally quiescent
galaxies with bulge dominated morphology \citep{kauffmann03a}. The
more massive members in these minor pairs may curtail their star
formation through mass quenching \citep{binney04, birnboim03, dekel06,
  das21}. They can also quench the star formation in their less
massive companions by stripping away gas and leading them to
starvation. Our analysis shows that the lighter members in minor pairs
experience a greater reduction in their star formation. The minor
interactions may thus enhance or suppress star formation in galaxy
pairs depending on their intrinsic luminosity. Our results suggest
that the minor interactions can significantly contribute to the
observed bimodality.

We have selected the control galaxies by closely matching the
redshift, stellar mass and environment that significantly reduces the
biases in our result. However, a few caveats remain in our
analysis. All the galaxies in close pairs may not be undergoing
interactions. Some of them may be approaching each other and are yet
to experience an encounter. There may also be pairs that are
undergoing interactions without any SFR enhancement. Further, there
are uncertainties in identifying galaxy pairs based on their projected
separation and velocity difference. Some of the selected pairs may not
be close in three dimensions. The chance superposition in the
high-density regions like groups and clusters may also produce
spurious pairs at both close and wide separations \citep{alonso04}.

Finally, we conclude that the galaxy interactions and environment both
play significant roles in galaxy evolution. The tidal interactions may
or may not trigger starbursts depending on the physical properties of
galaxies and their environment. We find that for the intrinsically
brighter galaxies, the major interactions enhance the star formation
rate in the low-density environments and suppress it in the high
density environments. On the other hand, the minor interactions in the
intrinsically brighter galaxies always quench star formation
irrespective of their environments. It would also be interesting to
study the effects of galaxy interactions in different geometric
environments of the cosmic web. We plan to carry out such an analysis
in future.

% add few references and revise texts

\section*{ACKNOWLEDGEMENT}

The authors thank the SDSS team for making the data publicly
available. BP would like to acknowledge financial support from the
SERB, DST, Government of India through the project CRG/2019/001110. BP
would also like to acknowledge IUCAA, Pune for providing support
through associateship programme. SS acknowledges IISER Tirupati for
support through a postdoctoral fellowship.

Funding for the SDSS and SDSS-II has been provided by the Alfred
P. Sloan Foundation, the Participating Institutions, the National
Science Foundation, the U.S. Department of Energy, the National
Aeronautics and Space Administration, the Japanese Monbukagakusho, the
Max Planck Society, and the Higher Education Funding Council for
England. The SDSS website is http://www.sdss.org/.

The SDSS is managed by the Astrophysical Research Consortium for the
Participating Institutions. The Participating Institutions are the
American Museum of Natural History, Astrophysical Institute Potsdam,
University of Basel, University of Cambridge, Case Western Reserve
University, University of Chicago, Drexel University, Fermilab, the
Institute for Advanced Study, the Japan Participation Group, Johns
Hopkins University, the Joint Institute for Nuclear Astrophysics, the
Kavli Institute for Particle Astrophysics and Cosmology, the Korean
Scientist Group, the Chinese Academy of Sciences (LAMOST), Los Alamos
National Laboratory, the Max-Planck-Institute for Astronomy (MPIA),
the Max-Planck-Institute for Astrophysics (MPA), New Mexico State
University, Ohio State University, University of Pittsburgh,
University of Portsmouth, Princeton University, the United States
Naval Observatory, and the University of Washington.

\end{document}